\documentclass[sigconf]{acmart}
\usepackage{algorithm}
\usepackage{multirow} 
\usepackage{enumitem}
\usepackage{xspace}
\usepackage{amsmath,amsfonts}
\usepackage{algorithmic}
\usepackage{acmart-taps}

\newcommand{\x}{$\times$\xspace}
\newcommand{\methodname}{AutoRAC\xspace}

\copyrightyear{2025}
\acmYear{2025}
\setcopyright{cc}
\setcctype{by}
\acmConference[GLSVLSI '25]{Great Lakes Symposium on VLSI 2025}{June 30-July 2, 2025}{New Orleans, LA, USA}
\acmBooktitle{Great Lakes Symposium on VLSI 2025 (GLSVLSI '25), June 30-July 2, 2025, New Orleans, LA, USA}
\acmDOI{10.1145/3716368.3735229}
\acmISBN{979-8-4007-1496-2/2025/06}

\begin{document}

\title{AutoRAC: Automated Processing-in-Memory Accelerator Design for Recommender Systems}

\author{Feng Cheng}
\affiliation{
  \institution{Duke University}
  \city{Durham}
  \state{NC}
  \country{USA}
}
\email{feng.cheng@duke.edu}

\author{Tunhou Zhang}
\affiliation{
  \institution{Duke University}
  \city{Durham}
  \state{NC}
  \country{USA}
}
\email{tunhou.zhang@duke.edu}

\author{Junyao Zhang}
\affiliation{
  \institution{Duke University}
  \city{Durham}
  \state{NC}
  \country{USA}
}
\email{junyao.zhang@duke.edu}

\author{Jonathan Hao-Cheng Ku}
\affiliation{
  \institution{Duke University}
  \city{Durham}
  \state{NC}
  \country{USA}
}
\email{jonathan.ku@duke.edu}

\author{Yitu Wang}
\affiliation{
  \institution{Duke University}
  \city{Durham}
  \state{NC}
  \country{USA}
}
\email{yitu.wang@duke.edu}

\author{Xiaoxuan Yang}
\affiliation{
  \institution{University of Virginia}
  \city{Charlottesville}
  \state{VA}
  \country{USA}
}
\email{xiaoxuan@virginia.edu}

\author{Hai ``Helen'' Li}
\affiliation{
  \institution{Duke University}
  \city{Durham}
  \state{NC}
  \country{USA}
}
\email{hai.li@duke.edu}

\author{Yiran Chen}
\affiliation{
  \institution{Duke University}
  \city{Durham}
  \state{NC}
  \country{USA}
}
\email{Yiran.chen@duke.edu}

\renewcommand{\shortauthors}{Cheng, et al.}


\begin{abstract}

The performance bottleneck of deep-learning-based recommender systems resides in their backbone Deep Neural Networks. 
By integrating Processing-In-Memory~(PIM) architectures, researchers can reduce data movement and enhance energy efficiency, paving the way for next-generation recommender models. 
Nevertheless, achieving performance and efficiency gains is challenging due to the complexity of the PIM design space and the intricate mapping of operators. 
In this paper, we demonstrate that automated PIM design is feasible even within the most demanding recommender model design space, spanning over $10^{54}$ possible architectures.
We propose \methodname, which formulates the co-optimization of recommender models and PIM design as a combinatorial search over mixed-precision interaction operations, and parameterizes the search with a one-shot supernet encompassing all mixed-precision options. 
We comprehensively evaluate our approach on three Click-Through Rate benchmarks, showcasing the superiority of our automated design methodology over manual approaches. 
Our results indicate up to a 3.36$\times$ speedup, 1.68$\times$ area reduction, and 12.48$\times$ higher power efficiency compared to naively mapped searched designs and state-of-the-art handcrafted designs.

\end{abstract}



\maketitle

\section{Introduction}

Advances in recommender systems have focused on enhancing personalization, scalability, and diversity. 
The incorporation of deep learning techniques~\cite{cheng2016wide, naumov2019deep, wang2021dcn, sheng2021one} has significantly improved the ability to capture and anticipate user preferences, enabling more accurate and tailored recommendations. 
Processing-in-Memory~(PIM) architectures~\cite{retrans, sramimc, bwq} offer promising pathways for next generation recommender models. 
First, PIM embeds computation within memory units, reducing data movement and improving energy efficiency. 
Second, PIM leverages emerging memory technologies, such as Resistive Random-Access Memory~(ReRAM) and Phase-Change Memory~(PCM), which provide higher density and lower latency for large datasets. 
Consequently, PIM-based solutions are compelling candidates for addressing the challenges faced by modern recommender systems.

\begin{figure*}[ht!]
    \centering
    \includegraphics[width=\textwidth]{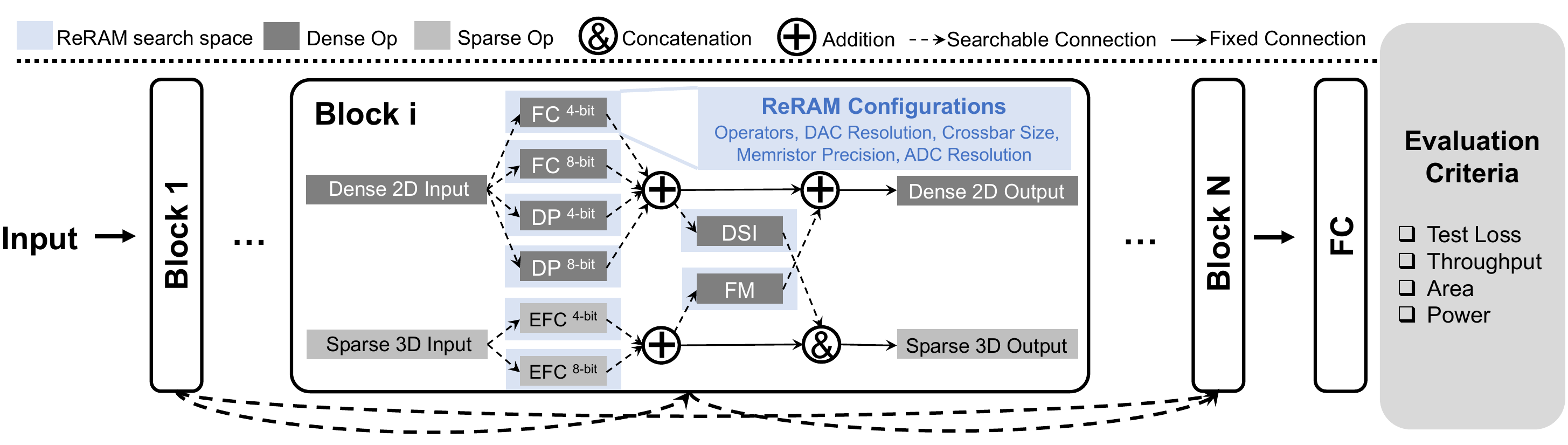} 
    \vspace{-2em}
    \caption{Overview of \methodname framework with search space and evaluation criteria.}
    \Description{Overview of \methodname framework with search space and evaluation criteria.}
    \label{fig:ss}
    \vspace{-1em}
\end{figure*}

Nevertheless, we posit that jointly optimizing the recommender system model and the PIM design can yield substantial improvements in system-level performance and efficiency. 
Our objective is to discover a PIM-friendly recommender model and to devise a dedicated PIM architecture for high-throughput, efficient inference. 
From the model-search perspective, a thorough exploration of a recommender system should encompass its operators, connections, embedding dimensions, and feature dimensions. 
From the PIM-design perspective, a chief obstacle arises from the unfixed operands in dot-product layers and factorization machine operations~\cite{guo2017deepfm}, necessitating additional crossbar programming steps during inference. 
Moreover, modifications to connections and operators during the search phase can significantly influence dataflow and mapping efficiency. 
Finally, the precision requirements of recommender models make them sensitive to nonidealities in certain analog PIM implementations~\cite{yang2021multi}.

The rise of Automated Machine Learning~(AutoML), particularly Neural Architecture Search~(NAS), has driven substantial innovations in designing recommender models by enabling algorithm-level optimizations~\cite{liu2018darts, yu2020bignas} and hardware-aware adaptations~\cite{lin2021naas}. 
While most NAS algorithms have been evaluated on relatively small-scale vision and language benchmarks, adapting NAS to recommender systems introduces unique challenges, especially when co-designing backbone architectures (e.g., DNNs) and specialized hardware (e.g., PIM). 
Two factors exacerbate these difficulties: 
first, recommender models require rigorous evaluation protocols, wherein even a minor shift (e.g., 0.2\% in Log Loss, or 0.001) can be critical; 
second, the co-design of deep neural network architectures and PIM hardware for recommender systems remains underexplored, with specialized operations (e.g., factorization machines) requiring dedicated consideration. 
These issues highlight the necessity of a NAS approach tailored to unlock the potential of joint optimization for PIM-based recommender models.

In this paper, we present a framework called \methodname, which employs NAS to accelerate recommender systems on PIM. 
We construct a comprehensive design space that encompasses recommender models and PIM systems, parameterized through a one-shot supernet covering all mixed-precision options.
To navigate this complex space, we adopt an evolutionary algorithm. 
Additionally, we streamline the search space to be hardware-friendly and aligned with PIM dataflow, thereby promoting both efficiency and efficacy in the search process. 
Our empirical findings showcase the advantages of our NAS-based design strategy and underscore the importance of well-structured dataflow. 
More crucially, our work provides valuable insights into the interplay among neural architectures and hardware configurations in PIM-based recommender systems. We make the following contributions in this paper.

\aptLtoX[graphic=no,type=html]{
\begin{itemize}[noitemsep,leftmargin=*]
    \item We propose \methodname, a holistic methodology for co-optimizing recommender-system architectures and PIM hardware, aiming to enhance overall system performance.
    \item We introduce a wide-ranging design space that spans over $10^{54}$ possible architectures, incorporating model structure, quantization, and PIM design, thereby demonstrating the feasibility of automated PIM design under demanding conditions.
    \item We present novel mapping schemes for operations such as dot product and factorization machine, accompanied by a carefully designed pipeline to manage their interconnections effectively.
    \item We extensively evaluate our proposed method on three CTR benchmarks, revealing that automated co-design achieves higher throughput, smaller area footprint, and better power efficiency compared to manual design.

\end{itemize}} {\setlist{nolistsep}
\begin{itemize}[leftmargin=1.2em]
    \item We propose \methodname, a holistic methodology for co-optimizing recommender-system architectures and PIM hardware, aiming to enhance overall system performance.
    \item We introduce a wide-ranging design space that spans over $10^{54}$ possible architectures, incorporating model structure, quantization, and PIM design, thereby demonstrating the feasibility of automated PIM design under demanding conditions.
    \item We present novel mapping schemes for operations such as dot product and factorization machine, accompanied by a carefully designed pipeline to manage their interconnections effectively.
    \item We extensively evaluate our proposed method on three CTR benchmarks, revealing that automated co-design achieves higher throughput, smaller area footprint, and better power efficiency compared to manual design.
\end{itemize}}

\section{Background and Related Work}

\noindent \textbf{Deep-Learning Recommender Systems.}
The remarkable success of deep learning has led to the broad adoption of DNNs~\cite{cheng2016wide,naumov2019deep,wang2021dcn} over traditional recommender designs~\cite{rendle2012factorization,he2017neural} for tasks such as Click-Through Rate~(CTR) prediction. 
Recently, NAS~\cite{yu2020bignas,liu2018darts} has emerged as a powerful methodology for algorithm-level optimization in recommender systems, contributing to end-to-end DNN architecture design~\cite{song2020towards,zhang2023NASRec}, feature-interaction selection~\cite{liu2020autofis,zhang2023distdnas}, and embedding-table optimization~\cite{zhaok2021autoemb}.
However, these approaches do not fully exploit hardware-aware optimizations, overlooking potential efficiency gains that arise from co-designing models and hardware. 
\methodname addresses this gap by simultaneously exploring DNN backbones and PIM hardware to provide holistic and practical solutions for recommender systems.

\noindent \textbf{PIM Designs.}
PIM architectures harness crossbar-based structures in memory technologies such as ReRAM~\cite{yang2022research}. As shown in Figure~\ref{fig:xb}a, analog voltages are applied to word lines (WLs) and multiplied by the conductances along each row (Ohm's Law). The resulting currents are then summed along each column (Kirchhoff's Current Law) and read out by circuitry connected to bit lines (BLs). These crossbar arrays thus naturally support matrix-vector multiplication~(MVM)~\cite{hu2014memristor}. 
Prior efforts have leveraged PIM parallelism to accelerate recommender systems, achieving promising gains~\cite{yang2023pimpr,wang2021rerec}. 
Nonetheless, these works do not address the unique challenges of PIM-based recommender systems, including inefficient hand-crafted mapping and heuristic-driven hardware design. Consequently, integrating PIM design into a unified search space is vital for delivering end-to-end solutions. 
\methodname tackles this need by constructing an optimized PIM-based recommender system, complete with improved processing engines and automated mapping strategies, ultimately identifying optimal architectures based on defined search criteria.
\section{\methodname}
In this section, we introduce \methodname, a unified framework designed to jointly optimize DNN backbones and PIM hardware for recommender systems. 
Figure~\ref{fig:ss} presents an overview of the \methodname workflow. 
We first detail the \methodname design space in Section~\ref{sec:design_space}, discussing both the DNN backbone and the PIM architecture search. 
Next, Section~\ref{sec:op_mapping} describes how we map DNN operators to PIM hardware under various design configurations. 
Subsequently, Section~\ref{sec:arch_overview} outlines the composition of the PIM-based recommender system architecture, which emerges from the combined optimization of the DNN design space and operator mappings. 
Finally, Section~\ref{sec:evo_search} elaborates on the automated evolutionary search process that drives the co-design approach in \methodname.

\vspace{-1em}

\subsection{\methodname Design Space} \label{sec:design_space}
To achieve a holistic co-optimization strategy for recommender systems, \methodname expands on two primary axes: the DNN backbone design space and the PIM design space. 
Table~\ref{tab:ss_conf} provides an overview of the configuration parameters across these two domains. 
Within the DNN design space, we target the selection of operators and interconnections vital for accurate and efficient recommendation. 
Within the PIM design space, we explore quantization techniques and ReRAM parameters that exploit the parallelism of PIM accelerators and effectively control hardware overhead. 
By encompassing a broad set of reasonable configurations while maintaining search tractability, \methodname increases the chance of discovering architectures that excel in both accuracy and efficiency.

\begin{table}[t]
\centering
\caption{\methodname design space construction}
\vspace{-0.8em}
\small
\begin{tabular}{c|c}
    \toprule
    \multicolumn{2}{c}{Model design space} \\
    \midrule
    Operator & FC, EFC, DP, DSI, FM \\
    Connection & Block-wise, Operator-wise \\
    Dense Feature Dimension & 16, 32, 64, 128, 256, 512, 768, 1024 \\
    Sparse Feature Dimension & 16, 32, 48, 64 \\
    \bottomrule
    \toprule
    \multicolumn{2}{c}{Quantization design space} \\
    \midrule
    Weight Quantization & 4, 8 \\
    \bottomrule
    \toprule
    \multicolumn{2}{c}{ReRAM design space} \\
    \midrule
    DAC Resolution & 1, 2 \\
    Crossbar Size & 16, 32, 64 \\
    Memristor Precision & 1, 2 \\
    ADC Resolution & 4, 6, 8 \\
    \bottomrule
\end{tabular}
\label{tab:ss_conf}
\end{table}

\noindent \textbf{Recommender Model Design Space.}
We adopt a design space inspired by NASRec~\cite{zhang2023NASRec}, adapting it for PIM-oriented dataflows. The model is composed of $N$ choice blocks followed by a final Fully-Connected~(FC) layer. Each choice block ingests an arbitrary number of dense tensors, $X_d \in \mathbb{R}^{B \times \textit{dim}_d}$, and sparse tensors, $X_s \in \mathbb{R}^{B \times N_s \times \textit{dim}_s}$, producing one dense output $Y_d$ and one sparse output $Y_s$. The operators are categorized as follows:
\aptLtoX[graphic=no,type=html]{\begin{itemize}
    \item \textit{Dense operators}, such as FC and Dot-Product~(DP), which output dense tensors.
    \item \textit{Sparse operators}, for instance, Embedded Fully-Connected~(EFC), which preserve sparse output structure.
    \item \textit{Dense--Sparse interaction operators}, which fuse information across dense and sparse branches. For example, a Dense-to-Sparse Merger (DSI) employs FC and reshaping to merge dense outputs into sparse features, while a Factorization Machine~(FM) acts as a Sparse-to-Dense Merger.
\end{itemize}}{\setlist{nolistsep}
\begin{itemize}[leftmargin=1.2em]
    \item \textit{Dense operators}, such as FC and Dot-Product~(DP), which output dense tensors.
    \item \textit{Sparse operators}, for instance, Embedded Fully-Connected~(EFC), which preserve sparse output structure.
    \item \textit{Dense--Sparse interaction operators}, which fuse information across dense and sparse branches. For example, a Dense-to-Sparse Merger (DSI) employs FC and reshaping to merge dense outputs into sparse features, while a Factorization Machine~(FM) acts as a Sparse-to-Dense Merger.
\end{itemize}}

We permit flexible connections between blocks, subject to the constraint that at least one operator is selected in the dense branch and one operator is selected in the sparse branch. 
This design ensures that a broad range of architectures, with differing operator orders and topological connections, is included in the search space.

\begin{figure}[t]
  \centering
  \includegraphics[width=\linewidth]{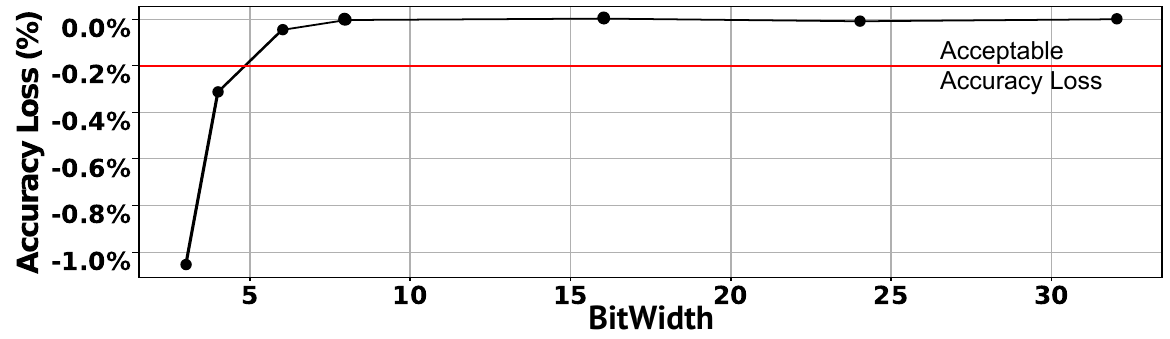}
  \vspace{-2em}
  \caption{Test Log Loss on Criteo versus weight bit-width.}
  \Description{Test Log Loss on Criteo versus weight bit-width.}
  \label{fig:ctr_criteo_loss}
  \vspace{-1em}
\end{figure}

\noindent \textbf{PIM Design Space.} The PIM design space covers two main aspects: quantization and ReRAM configurations, both of which are crucial to maximizing speedup and energy efficiency in recommender systems.

\aptLtoX[graphic=no,type=html]{\begin{itemize}
\item \textit{Quantization Design Space.} We begin with a 32-bit floating-point representation for both weights and activations, then progressively reduce bit-width. 
    Empirical testing on the Criteo dataset reveals that accuracy remains relatively stable at higher precisions, but begins to degrade sharply if weights drop below 8 bits (see Figure~\ref{fig:ctr_criteo_loss}). 
    Moreover, lowering activation bit-width further complicates convergence for the supernet in large-scale recommendation tasks. 
    To keep the design space tractable, \methodname restricts weight precision to 4~bits or 8~bits for FC, EFC, DSI, DP and FM operators. This choice strikes a balance between hardware efficiency and model fidelity. 
    We exclude 6-bit weight quantization because it is not a power-of-two format, which, according to our experiments, reduces crossbar utilization and tends to hinder overall performance.
    \item \textit{ReRAM Design Space.} We tailor the ReRAM design space to meet the stringent low-loss requirements typically required by recommender systems~\cite{li2022autolossgen}. 
    The configuration of this space is intuited by previous research~\cite{yang2021multi}, with adjustments to minimize the impact of the intrinsic nonidealities of ReRAM crossbar arrays. Options for the crossbar size include 16, 32, and 64, while memristor precision and Digital-to-Analog Converter (DAC) resolution are set to 1 or 2 bits. For the Analog-to-Digital Converter (ADC), precision options are 4, 6 and 8 bits. Notably, we only consider combinations of DAC and memristor precision that fall within the maximum ADC resolution range to avoid any loss during the analog-to-digital conversion process. 
    Although this constraint may slightly reduce design space, it is a deliberate choice to ensure that the resulting models exhibit lower loss.

\end{itemize}}{\setlist{nolistsep}
\begin{itemize}[leftmargin=1.2em]
    \item \textit{Quantization Design Space.} We begin with a 32-bit floating-point representation for both weights and activations, then progressively reduce bit-width. 
    Empirical testing on the Criteo dataset reveals that accuracy remains relatively stable at higher precisions, but begins to degrade sharply if weights drop below 8 bits (see Figure~\ref{fig:ctr_criteo_loss}). 
    Moreover, lowering activation bit-width further complicates convergence for the supernet in large-scale recommendation tasks. 
    To keep the design space tractable, \methodname restricts weight precision to 4~bits or 8~bits for FC, EFC, DSI, DP and FM operators. This choice strikes a balance between hardware efficiency and model fidelity. 
    We exclude 6-bit weight quantization because it is not a power-of-two format, which, according to our experiments, reduces crossbar utilization and tends to hinder overall performance.
    \item \textit{ReRAM Design Space.} We tailor the ReRAM design space to meet the stringent low-loss requirements typically required by recommender systems~\cite{li2022autolossgen}. 
    The configuration of this space is intuited by previous research~\cite{yang2021multi}, with adjustments to minimize the impact of the intrinsic nonidealities of ReRAM crossbar arrays. Options for the crossbar size include 16, 32, and 64, while memristor precision and Digital-to-Analog Converter (DAC) resolution are set to 1 or 2 bits. For the Analog-to-Digital Converter (ADC), precision options are 4, 6 and 8 bits. Notably, we only consider combinations of DAC and memristor precision that fall within the maximum ADC resolution range to avoid any loss during the analog-to-digital conversion process. 
    Although this constraint may slightly reduce design space, it is a deliberate choice to ensure that the resulting models exhibit lower loss.
\end{itemize}}

The design space is summarized in Table \ref{tab:ss_conf}. We include the model design space, quantization design space, and ReRAM design space in the table. To simplify the search process, we fixed the number of searchable blocks to \(N = 7\), encompassing as many as 2 \x $10^{54}$ architectures characterized by significant heterogeneity. Due to the limited use of human-derived priors and this vast, unrestricted search space, exhaustive sampling-based approaches could require an extensive amount of time to identify a cutting-edge model.

\begin{figure}[t]
  \centering
  \includegraphics[width=\linewidth]{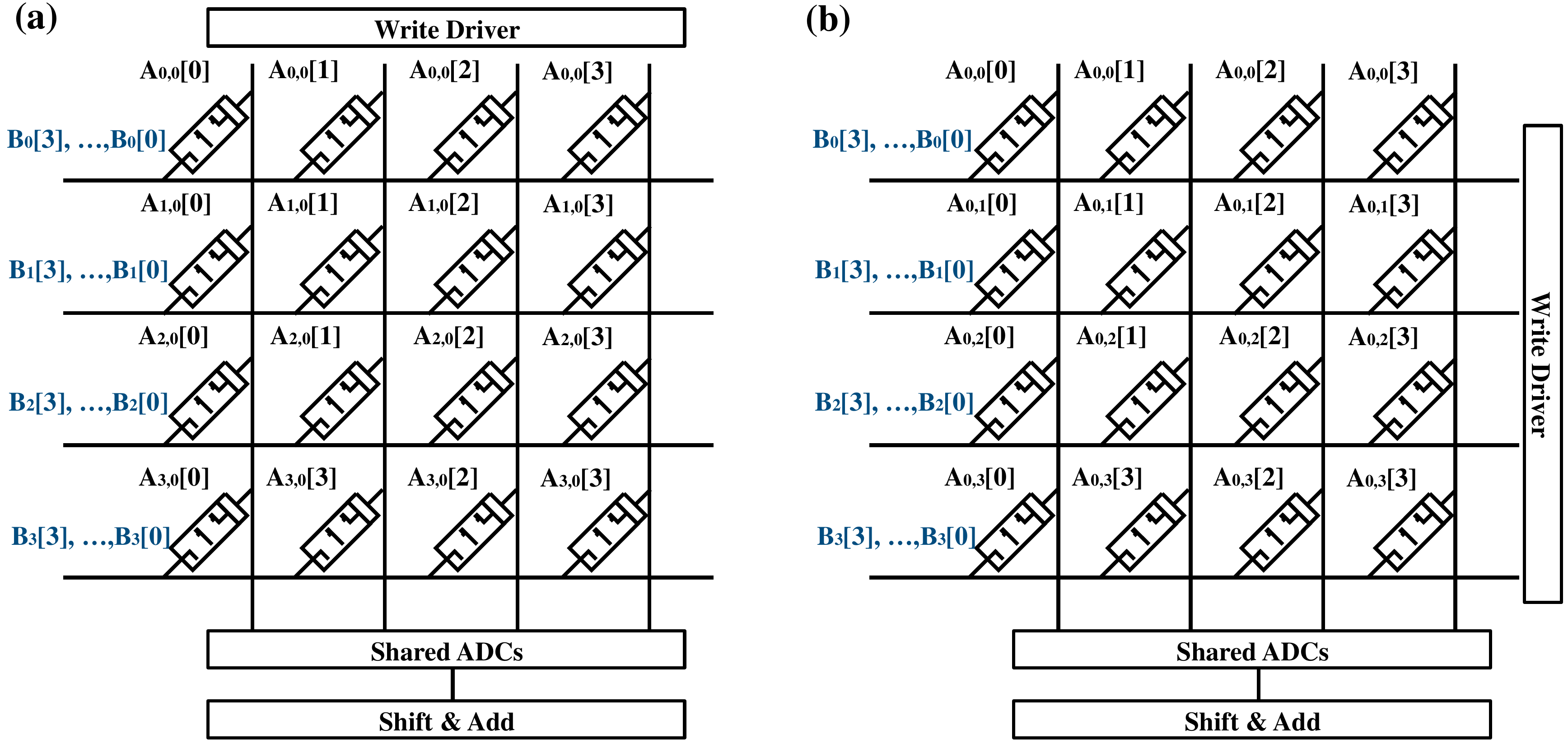}
  \caption{(a) ReRAM crossbar for FC, EFC, and DP. (b) Transposed-write ReRAM crossbar for FM.}
  \Description{(a) ReRAM crossbar for FC, EFC, and DP. (b) Transposed-write ReRAM crossbar for FM.}
  \label{fig:xb}
  \vspace{-1em}
\end{figure}

\subsection{Operators Mapping} \label{sec:op_mapping}
In this section, we illustrate the mapping between DNN operators and PIM hardware to design PIM-based recommender models. For ease of explanation, we set the batch size to one.

\begin{figure*}[ht!]
    \centering
    \includegraphics[width=\textwidth]{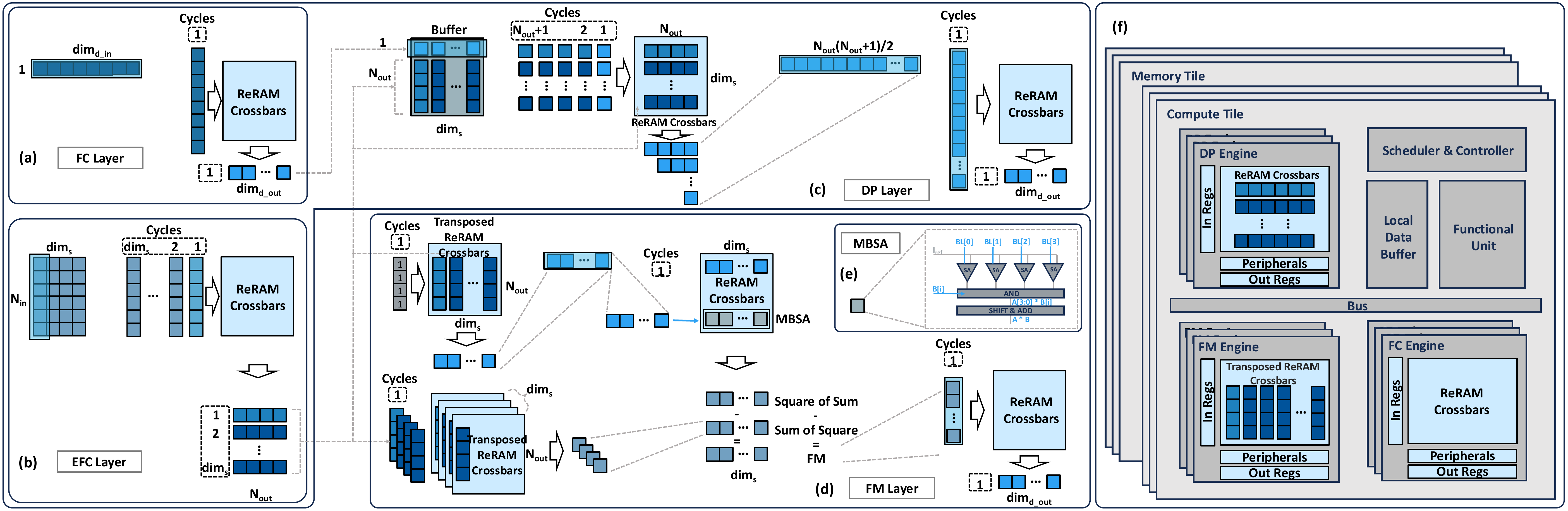}
    \vspace{-10pt}
    \caption{Overview of \methodname mapping schemes and architecture design.}
    \Description{Overview of \methodname mapping schemes and architecture design.}
    \label{fig:mapping}
    \vspace{-1em}
\end{figure*}

\noindent \textbf{FC Layer, EFC Layer, and DSI. } 
FC layers and EFC layers are pivotal for generating dense representations and constructing sparse features in recommender systems. 
The DSI is functionally an FC layer followed by a reshaping operation, enabling the transition of feature representations from a dense format to a sparse one. 
In contrast, an EFC layer performs sparse computations along a middle dimension as \(Y_s = W_s X_s,\)
where \(W_s \in \mathbb{R}^{\text{N}_{in} \times \text{N}_{out}}\) is the weight matrix, \(X_s \in \mathbb{R}^{\text{N}_{in} \times \text{dim}_s}\) is the input tensor, and \(Y_s \in \mathbb{R}^{\text{N}_{out} \times \text{dim}_s}\) is the output tensor. 
Both FC and EFC layers essentially perform vector--matrix or matrix--matrix multiplications, which map naturally onto ReRAM crossbars. 
As illustrated in Figures~\ref{fig:mapping}a and~\ref{fig:mapping}b, these layers are implemented in PIM by programming the weight matrix~\(W\) onto the crossbar and sequentially feeding the bits of input vector~\(X\) to the word lines. 
The ReRAM crossbars then perform MVMs, producing the output tensor~\(Y\). 
For simplicity, Figure~\ref{fig:mapping} omits the explicit visualization of individual bits.

\noindent \textbf{DP Layer.} The DP layer captures feature interactions by computing pairwise inner products across multi-modal inputs, accommodating both dense and sparse feature representations. Its design involves four components: an FC layer for mapping the dense dimension (\(dim_d\)) to the sparse dimension (\(dim_s\)), an EFC layer for projecting the number of sparse features (\(N_s\)) to \(\sqrt{2 \times dim_d}\), a dedicated DP engine for inner products, and a final FC layer for projecting the concatenated result to the desired output dimension. 

Initially, the dense input is reshaped via an FC layer to match the sparse dimension, and the sparse features are simultaneously reduced through an EFC layer to \(\sqrt{2 \times dim_d}\) for balanced operator workloads. The outputs of these two transformations are merged into a single tensor 
\(\,X \in \mathbb{R}^{(\sqrt{2 \times dim_d}+1) \times dim_s}\), 
which undergoes pairwise inner products computed as 
\(\,\text{Triu}(XX^T)\).
The flattened inner-product results are then passed through an FC layer, yielding the final output.

Mapping of the FC and EFC submodules onto ReRAM crossbars remains consistent with the approach outlined for FC/EFC layers. 
Figure~\ref{fig:mapping}c highlights the additional steps required for the DP engine. 
Specifically, each output vector from the EFC and FC layers is buffered and programmed onto crossbars. Meanwhile, the EFC layer generates the next vector output, enabling a pipelined, overlap-friendly process that avoids unnecessary waiting. 
Because the sparse output from the EFC layer is inherently transposed, \(X^T\) can be programmed directly. Once the sparse feature matrix is fully produced, each feature vector is loaded into the crossbars to compute partial dot-product results, which are concatenated into the DP layer's final output. 
This output subsequently flows into an FC layer, which projects it to the final dense dimension \(dim_{\text{d\_out}}\).

\noindent \textbf{Sparse-to-Dense FM Layer.} 
The FM layer consists of two components: an FM engine for converting a 3D sparse representation into a dense vector, followed by an fully-connected layer for mapping this dense vector to the desired output size. 
The FM engine processes a batch of three-dimensional sparse tensors by computing 
\(\left( \sum_{i=1}^{n} \mathbf{x}_i \right)^2\) 
and 
\(\sum_{i=1}^{n} \mathbf{x}_i^2\). 
Here, \(\mathbf{x}_i\) denotes each embedding in the sparse feature set. The interaction term 
\(\,\mathbf{ix} = \left(\sum_{i=1}^{n} \mathbf{x}_i\right)^2 - \sum_{i=1}^{n} \mathbf{x}_i^2\,\) 
results from subtracting the sum of squares from the square of the sum, effectively capturing pairwise feature interactions.

We propose a novel PIM mapping for the FM operator, focusing on the square of the sum and the sum of the squares. 
Figure~\ref{fig:mapping}d provides an overview of this process. 
To calculate the square of the sum, the outputs of the EFC layer are programmed into the columns of a transposed ReRAM array~\cite{transpose_xb} (Figure~\ref{fig:xb}b). 
Unlike traditional crossbar architectures that program sparse outputs row by row, this transposed layout aligns spatially with the inputs and eliminates idle buffers. 
Once all sparse outputs are programmed, a vector of ones is supplied to the word lines to accumulate each column’s sum. 
Next, element-wise multiplication is performed on these sum vectors using the MBSA~\cite{mbsa} module (Figure~\ref{fig:mapping}e). 
First, the sum vector is programmed; then each bit is sent in parallel to the MBSA’s AND gates. 
Iterating this process across every bit ultimately produces the square of the sum.

The sum of squares is computed in parallel by directly programming each vector output of the EFC layer onto the transposed crossbar. 
Because each word line receives an identical vector, each row of the crossbar naturally yields a squared value. 
These values are then summed along the bit lines, delivering the aggregate sum of squares. 
Critically, the operations for the sum of squares and the square of the sum can be performed concurrently, leveraging the transposed array’s capability for full data pipelining. 
Afterwards, the difference between these two computations is fed into the final FC layer implemented on the ReRAM crossbars, projecting the result to the designated output dimension. 
This integrated procedure expedites throughput and reduces latency, culminating in an efficient factorization mechanism on PIM.

\vspace{-1em}
\subsection{Architecture Overview} 
\label{sec:arch_overview}
We illustrate the overall system architecture in Figure~\ref{fig:mapping}f. 
This design integrates memory tiles for storing embedding tables alongside computation tiles for operator execution. 
The memory tiles hold embedding tables in a static, read-only state, and an offline access-aware mechanism reorganizes embeddings by their frequency of occurrence, placing them in round-robin fashion across different banks to avoid conflicts. 
Meanwhile, the computation tiles are partitioned into three dedicated engines: the FM engine, DP engine, and MVM engine. 
Each engine hosts a crossbar array with its peripheral circuitry and I/O registers, mirroring the PIM functionality detailed in the preceding sections. 
Additionally, each tile contains a data buffer for intermediate outputs and a functional unit responsible for activation functions. 
A controller and a scheduler coordinate the data flow, ensuring all pipeline stages run smoothly and efficiently.

\vspace{-1em}

\subsection{Evolutionary Search}
\label{sec:evo_search}
The step-by-step search procedure is outlined in Algorithm~\ref{algo:NeuroRAC}. 
We adopt a regularized-evolution framework to identify the optimal subnet configuration within the \methodname design space. 
On the model side, each iteration selects one parent configuration, then applies a series of actions within a chosen block, such as swapping dense/sparse operators, modifying dense/sparse dimensions, adjusting block-to-block connections, or introducing dense--sparse interaction layers. 
Concurrently, the PIM design space is explored using a similar evolutionary process, but with mutations specialized to toggling among different ADC resolutions, DAC options, memristor precisions, and crossbar sizes. 
This targeted mutation scheme is crucial for uncovering architectures that align with performance and efficiency requirements in PIM-based recommender systems. 
By systematically switching these hardware parameters, we can comprehensively evaluate and refine the PIM design under diverse computational demands and constraints.

\begin{algorithm}[b]     
\caption{Best Subnet Config Search in \methodname}
\label{algo:NeuroRAC}

\begin{algorithmic}[1]  
\REQUIRE Design targets $\bigl[\tfrac{1}{\text{throughput}},\;\text{area},\;\text{power}\bigr]$
         denoted by  $\bigl[\text{target}_1,\text{target}_2,\text{target}_3\bigr]$

\STATE $all\_populations \leftarrow \operatorname{random\_search}(supernet)$

\FOR{$generation \leftarrow 1 \;\textbf{to}\; num\_generations$}
    \STATE $parent \leftarrow \operatorname{Sample\_and\_select}(criterion,\, all\_populations)$

    \FOR{$child \leftarrow 1 \;\textbf{to}\; num\_children$}
        \STATE $choice \leftarrow parent$
        \FOR{$mutation \leftarrow 1 \;\textbf{to}\; num\_mutations$}
            \STATE $choice \leftarrow \operatorname{Mutate}(choice)$
        \ENDFOR

        \STATE $\text{test\_loss} \leftarrow \operatorname{finetune\_and\_eval\_loss}(choice)$
        \STATE $\text{metric} \leftarrow \operatorname{hw\_ea}(choice)$
        \STATE $criterion \leftarrow \text{test\_loss}
                + \sum_{i=1}^{3}\lambda_i
                  \frac{\text{metric}_i}{\text{target}_i}$

        \STATE append $(choice,\,criterion)$ to $all\_populations$
    \ENDFOR

    \STATE sort $all\_populations$ by $criterion$
    \STATE remove last $num\_children$ entries
\ENDFOR
\end{algorithmic}
\end{algorithm}

\section{Evaluation}

\subsection{Experiment setup and Benchmark}
\noindent \textbf{Experiment Setup.}
To model on-chip buffers, we use CACTI~\cite{cacti} at a 32\,nm technology node. 
For ReRAM characterization, we follow the parameters in MNSIM2.0~\cite{mnsim} to obtain precise estimates of area, latency, and power consumption. 
We develop a behavioral simulator to further analyze end-to-end latency and throughput. 
Although our primary exploration and performance simulations are executed on an Intel Xeon Gold 6254 platform, we leverage an NVIDIA A5000 GPU to accelerate the co-exploration process.

\noindent \textbf{Recommender Model Benchmarks.}
We conduct empirical evaluations on three widely used CTR benchmarks: {Criteo}~\cite{criteo}, {Avazu}~\cite{avazu}, and {KDD Cup 2012}~\cite{kddcup}. 
The datasets are preprocessed following the same protocol used in NASRec~\cite{zhang2023NASRec}. 
After preprocessing, each dataset is split into a training set (80\%), a validation set (10\%), and a test set (10\%). 
During the \methodname~search phase, we train a supernet on the training set and identify the top 15 subnets based on validation performance. 
These 15 subnets are each retrained from scratch, and the best-performing one is selected as the final architecture.

\vspace{-1em}

\subsection{Experiment Results}
\noindent \textbf{Model Accuracy.}
Table~\ref{tab:ctr_results} compares \methodname with both hand-crafted and NAS-crafted baselines on three widely used CTR benchmarks: Criteo, Avazu, and KDD Cup 2012. Two central metrics, \emph{Log Loss} (lower is better) and \emph{AUC} (higher is better), are used to quantify predictive performance. 
On the Criteo dataset, \methodname achieves a Log Loss of 0.4397 and an AUC of 0.8116. Although these values may appear close to those of existing models, even a 0.001 reduction in Log Loss can yield notable gains in practical recommender systems. Notably, \methodname surpasses NASRec and outperforms hand-crafted approaches such as DLRM and DeepFM. 
For the Avazu dataset, \methodname sustains its strong results with a Log Loss of 0.3736 and an AUC of 0.7906, outperforming NASRec and highlighting its reliable predictive power across diverse user-item interactions. 
On the KDD Cup 2012 dataset, \methodname further validates its effectiveness by posting a Log Loss of 0.1489 and an AUC of 0.8160, ranking first in both metrics. These outcomes collectively underscore the model’s adaptability to varied data distributions and its capacity for capturing vital feature interactions through automated architecture search.

\begin{table}[h]
    \begin{center}
    \vspace{-1.25em}
    \caption{Performance of \methodname on CTR Tasks.}
    \vspace{-1.25em}
    \scalebox{0.7}{
    \begin{tabular}{|c|c|cc|cc|cc|c|}
    \hline
     & \multirow{2}{*}{\textbf{Method}} & \multicolumn{2}{|c|}{\textbf{Criteo}}  &  \multicolumn{2}{|c|}{\textbf{Avazu}} & \multicolumn{2}{|c|}{\textbf{KDD }}  \\
      & &  Log Loss & AUC & Log Loss & AUC & Log Loss & AUC  \\
    \hline \hline
    \multirow{4}{*}{\shortstack{\textbf{Hand}\\\textbf{crafted}}} & DLRM~\cite{naumov2019deep} & 0.4436 & 0.8085 & 0.3814 & 0.7766 & 0.1523 & 0.8004 \\
    & xDeepFM~\cite{lian2018xdeepfm} & 0.4418 & 0.8052 & - & - & - & - \\
    & AutoInt+~\cite{song2019autoint} & 0.4427 & 0.8090 & 0.3813 & 0.7772 & 0.1523 & 0.8002 \\
    & DeepFM~\cite{guo2017deepfm} & 0.4432 & 0.8086 & 0.3816 & 0.7767 & 0.1529 & 0.7974 \\
    \hline

    \multirow{2}{*}{\shortstack{\textbf{NAS}\\\textbf{crafted}}} & 
    NASRec~\cite{zhang2023nasrec2} & 0.4399 &\textbf{0.8118} & 0.3747 & 0.7887 & 0.1495 & 0.8135 \\
    & \methodname & \textbf{0.4397} &0.8116 & \textbf{0.3736} & \textbf{0.7906} & \textbf{0.1489} & \textbf{0.8160} \\
    \hline
    \end{tabular}
    \label{tab:ctr_results}
    }
    \end{center}
    \vspace{-1.25em}
\end{table}

\noindent \textbf{Hardware Performance}.
Table~\ref{tab:area} summarizes the hardware performance of \methodname in comparison with a CPU baseline, a naively mapped NASRec~\cite{zhang2023nasrec2} design, and two handcrafted accelerators, RecNMP~\cite{ke2019recnmp} and ReREC~\cite{wang2021rerec}. Three principal metrics are used in this comparison: speedup, power efficiency, and area savings. 
When measured against the CPU, \methodname achieves a 22.83$\times$ speedup while also improving power efficiency by 66.87$\times$, indicating that significant acceleration can be realized by leveraging specialized PIM hardware alongside the automatically discovered DNN architecture. This high degree of hardware–software co-optimization effectively reduces memory transfers and accelerates computations inherent in recommender systems. 
In comparison with the naively mapped NASRec, \methodname demonstrates a 3.17$\times$ speedup and achieves 2.39$\times$ higher power efficiency, complemented by a 1.68$\times$ reduction in area. These improvements highlight the importance of searching for hardware-friendly operator configurations in tandem with the model design. By reducing mismatches between dataflow patterns and physical crossbar layouts, \methodname avoids many of the inefficiencies encountered in naive mappings. 
The comparison against state-of-the-art handcrafted designs shows that \methodname remains highly competitive. In relation to RecNMP, \methodname attains a 12.48$\times$ improvement in power efficiency and a 3.36$\times$ speedup. Compared with ReREC, \methodname displays a 1.57$\times$ gain in power efficiency and a 1.28$\times$ speedup. The results underscore how a systematic, search-based approach can either match or exceed manually optimized accelerators by jointly refining both algorithmic and architectural choices.

\begin{table}[h]
    \begin{center}
    \vspace{-1.25em}
    \caption{Hardware metrics of \methodname against  baselines.}
    \small
    \vspace{-1.25em}
    \scalebox{0.8}{
    \begin{tabular}{|c|ccc|}
    \hline
         \methodname Against & Area Savings & Power Efficiency & Speedup \\
         \hline \hline
        CPU                            & -       & 66.87\x & 22.83\x \\ 
        RecNMP \cite{ke2019recnmp}     & -       & 12.48\x & 3.36\x \\
        NASRec \cite{zhang2023nasrec2}  & 1.68\x & 2.39\x & 3.17\x \\
        ReREC  \cite{wang2021rerec}    & -       & 1.57\x & 1.28\x \\
        \hline
    \end{tabular}
    \label{tab:area}
    }
    \end{center}
    \vspace{-1.25em}
\end{table}

\begin{figure}[h]
    \centering
    \vspace{-1em}
    \includegraphics[width=1.0\linewidth]{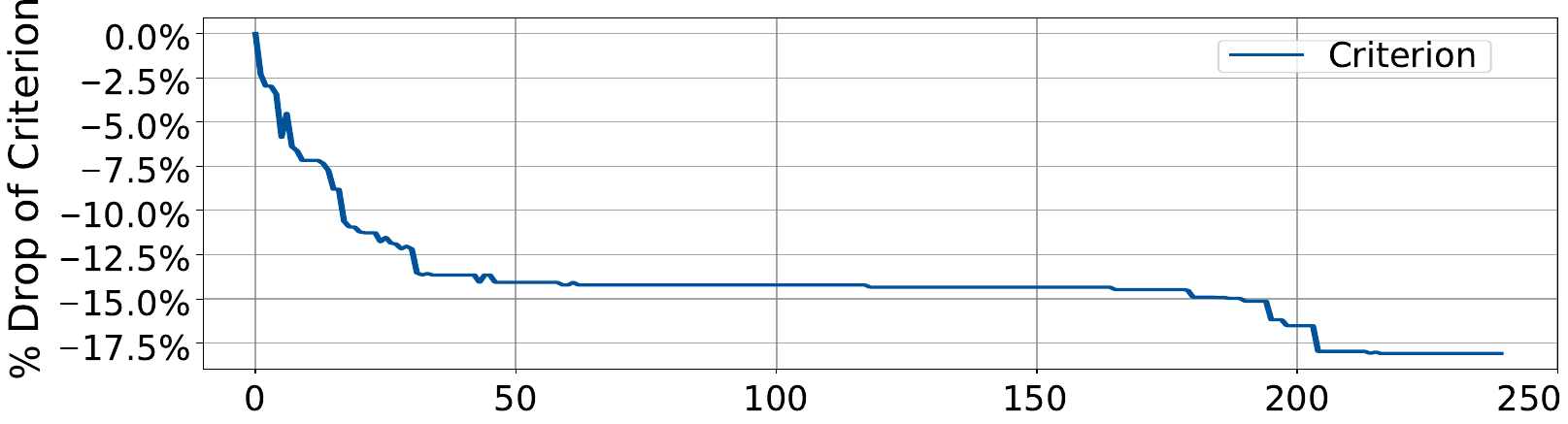} 
    \vspace{-2.5em}
    \caption{Percentage drop of criterion (Lower is better).}
    \Description{Percentage drop of criterion (Lower is better).}
    \label{fig:criterion}
    \vspace{-1em}
\end{figure}

\noindent \textbf{Search Efficiency}.
Figure~\ref{fig:criterion} illustrates the evolution of the performance criterion across 240 generations during the search. The criterion begins with a rapid decline of over 10\% within the first 50 generations, indicating that the search strategy quickly identifies promising model–hardware configurations and discards less suitable ones. After this initial period of rapid improvement, the curve plateaus, suggesting that the search converges to top-performing candidates, with only incremental gains observed. Around the 150th generation, another period of gradual performance increase emerges, indicating that although the search has discovered strong solutions, further exploration may reveal moderately better architectures. After 200 generations, the curve stabilizes and shows minimal further decrease, signifying that the algorithm has effectively exploited the design space to discover high-quality solutions.

\begin{figure}[h]
  \centering
  \vspace{-1em}
  \includegraphics[width=0.6\linewidth]{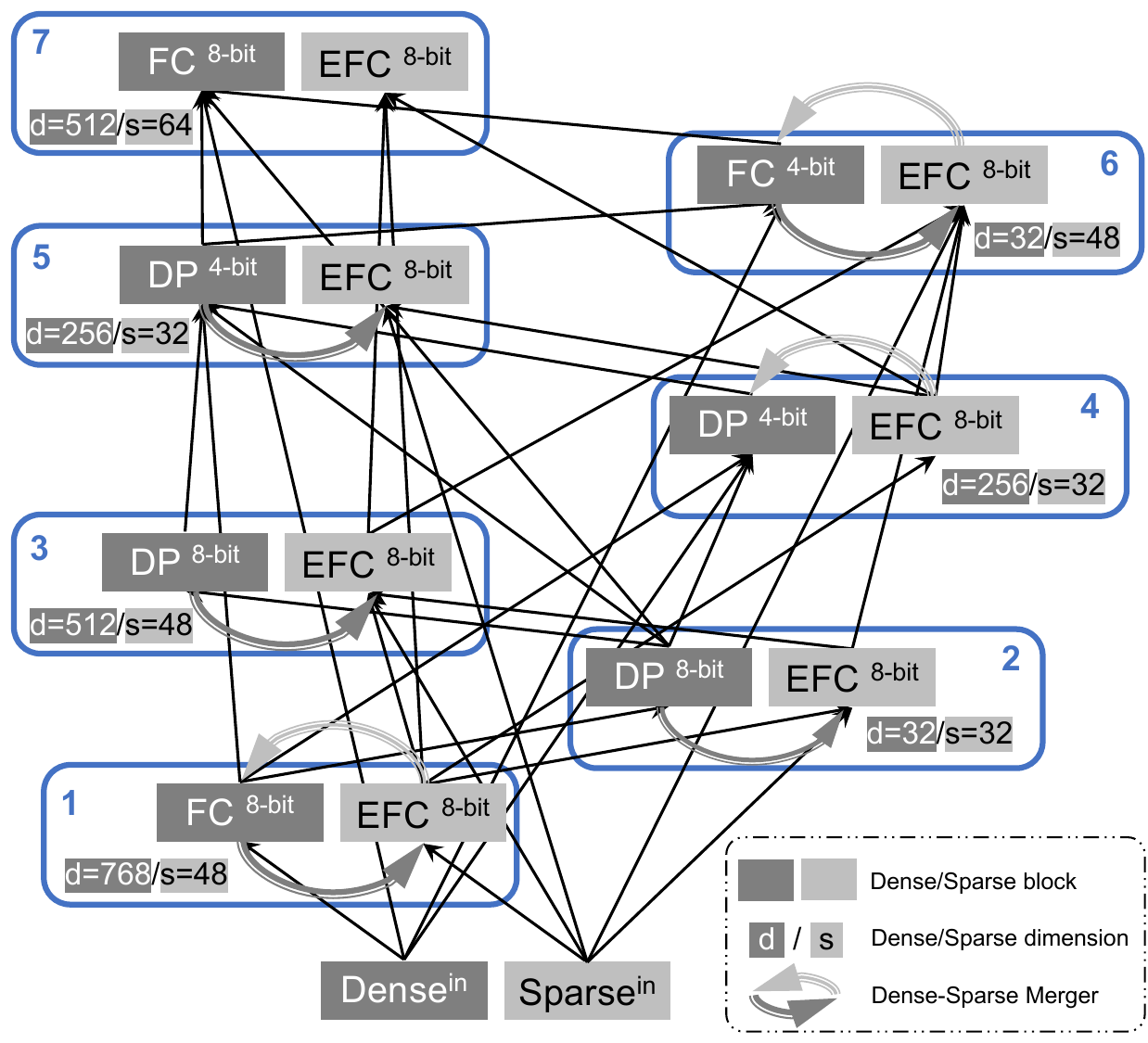}
  \vspace{-1em}
  \caption{Best model discovered from \methodname.}
  \Description{Best model discovered from \methodname.}
  \label{fig:best}
  \vspace{-2em}
\end{figure}

\noindent \textbf{Best Model Discovered}.
Figure~\ref{fig:best} depicts the best-performing architecture discovered on the Criteo dataset. This architecture reveals several noteworthy design trends. The EFC layers are predominantly 8-bit, a choice that likely results from their comparatively smaller parameter size, allowing them to maintain better precision without incurring an excessive resource burden. The FC layers in the middle of the network typically use 4-bit precision, an allocation that effectively balances computational overhead and accuracy in intermediate stages. In contrast, the initial and final FC layers generally adopt 8-bit precision, indicating that retaining more fine-grained details in the early and late phases of the network is beneficial for preserving critical information. The DP layers do not show a strong preference for any particular bit-width, suggesting that the design of these interaction-oriented modules can flexibly align with either higher or lower precision. Overall, the architecture discovered by \methodname strikes a nuanced balance between 4-bit and 8-bit operators, demonstrating the effectiveness of automatically searching for models that adapt precision settings to the computational needs of different layers.

\section{Conclusion}

This work demonstrates the practicality of automating processing-in-memory (PIM) design for large-scale recommender models. The proposed framework, \methodname, casts the joint optimization of DNN architectures and PIM hardware as a mixed-precision search over a one-shot supernet. Experimental results show improvements of up to 3.4\x in speed, 1.7\x reduction in silicon area, and 
12.5\x higher power efficiency compared with naïve mappings and state-of-the-art handcrafted baselines, underscoring the benefits of unified neural-architecture and hardware exploration.

\begin{acks}
The work was funded in part by National Science Foundation NSF-2328805 and NSF-2112562.
\end{acks}

\balance
\bibliographystyle{ACM-Reference-Format}
\bibliography{reference} 


\begin{thebibliography}{35}


\ifx \showCODEN    \undefined \def \showCODEN     #1{\unskip}     \fi
\ifx \showDOI      \undefined \def \showDOI       #1{#1}\fi
\ifx \showISBNx    \undefined \def \showISBNx     #1{\unskip}     \fi
\ifx \showISBNxiii \undefined \def \showISBNxiii  #1{\unskip}     \fi
\ifx \showISSN     \undefined \def \showISSN      #1{\unskip}     \fi
\ifx \showLCCN     \undefined \def \showLCCN      #1{\unskip}     \fi
\ifx \shownote     \undefined \def \shownote      #1{#1}          \fi
\ifx \showarticletitle \undefined \def \showarticletitle #1{#1}   \fi
\ifx \showURL      \undefined \def \showURL       {\relax}        \fi
\providecommand\bibfield[2]{#2}
\providecommand\bibinfo[2]{#2}
\providecommand\natexlab[1]{#1}
\providecommand\showeprint[2][]{arXiv:#2}

\bibitem[Balasubramonian et~al\mbox{.}(2017)]%
        {cacti}
\bibfield{author}{\bibinfo{person}{Rajeev Balasubramonian}, \bibinfo{person}{Andrew~B. Kahng}, \bibinfo{person}{Naveen Muralimanohar}, \bibinfo{person}{Ali Shafiee}, {and} \bibinfo{person}{Vaishnav Srinivas}.} \bibinfo{year}{2017}\natexlab{}.
\newblock \showarticletitle{{CACTI} 7: New Tools for Interconnect Exploration in Innovative Off-Chip Memories}.
\newblock \bibinfo{journal}{\emph{{ACM} Trans. Archit. Code Optim.}} \bibinfo{volume}{14}, \bibinfo{number}{2} (\bibinfo{year}{2017}), \bibinfo{pages}{14:1--14:25}.
\newblock


\bibitem[Cheng et~al\mbox{.}(2016)]%
        {cheng2016wide}
\bibfield{author}{\bibinfo{person}{Heng-Tze Cheng}, \bibinfo{person}{Levent Koc}, \bibinfo{person}{Jeremiah Harmsen}, \bibinfo{person}{Tal Shaked}, \bibinfo{person}{Tushar Chandra}, \bibinfo{person}{Hrishi Aradhye}, \bibinfo{person}{Glen Anderson}, \bibinfo{person}{Greg Corrado}, \bibinfo{person}{Wei Chai}, \bibinfo{person}{Mustafa Ispir}, \bibinfo{person}{Rohan Anil}, \bibinfo{person}{Zakaria Haque}, \bibinfo{person}{Lichan Hong}, \bibinfo{person}{Vihan Jain}, \bibinfo{person}{Xiaobing Liu}, {and} \bibinfo{person}{Hemal Shah}.} \bibinfo{year}{2016}\natexlab{}.
\newblock \showarticletitle{Wide \& Deep Learning for Recommender Systems}. In \bibinfo{booktitle}{\emph{Proceedings of the 1st Workshop on Deep Learning for Recommender Systems}} (Boston, MA, USA) \emph{(\bibinfo{series}{DLRS 2016})}. \bibinfo{publisher}{Association for Computing Machinery}, \bibinfo{address}{New York, NY, USA}, \bibinfo{pages}{7–10}.
\newblock
\showISBNx{9781450347952}
\urldef\tempurl%
\url{https://doi.org/10.1145/2988450.2988454}
\showDOI{\tempurl}


\bibitem[Guo et~al\mbox{.}(2017)]%
        {guo2017deepfm}
\bibfield{author}{\bibinfo{person}{Huifeng Guo}, \bibinfo{person}{Ruiming Tang}, \bibinfo{person}{Yunming Ye}, \bibinfo{person}{Zhenguo Li}, {and} \bibinfo{person}{Xiuqiang He}.} \bibinfo{year}{2017}\natexlab{}.
\newblock \showarticletitle{DeepFM: a factorization-machine based neural network for CTR prediction}. In \bibinfo{booktitle}{\emph{Proceedings of the 26th International Joint Conference on Artificial Intelligence}} (Melbourne, Australia) \emph{(\bibinfo{series}{IJCAI'17})}. \bibinfo{publisher}{AAAI Press}, \bibinfo{pages}{1725–1731}.
\newblock
\showISBNx{9780999241103}


\bibitem[He et~al\mbox{.}(2017)]%
        {he2017neural}
\bibfield{author}{\bibinfo{person}{Xiangnan He}, \bibinfo{person}{Lizi Liao}, \bibinfo{person}{Hanwang Zhang}, \bibinfo{person}{Liqiang Nie}, \bibinfo{person}{Xia Hu}, {and} \bibinfo{person}{Tat-Seng Chua}.} \bibinfo{year}{2017}\natexlab{}.
\newblock \showarticletitle{Neural Collaborative Filtering}. In \bibinfo{booktitle}{\emph{Proceedings of the 26th International Conference on World Wide Web}} (Perth, Australia) \emph{(\bibinfo{series}{WWW '17})}. \bibinfo{publisher}{International World Wide Web Conferences Steering Committee}, \bibinfo{address}{Republic and Canton of Geneva, CHE}, \bibinfo{pages}{173–182}.
\newblock
\showISBNx{9781450349130}
\urldef\tempurl%
\url{https://doi.org/10.1145/3038912.3052569}
\showDOI{\tempurl}


\bibitem[Hu et~al\mbox{.}(2014)]%
        {hu2014memristor}
\bibfield{author}{\bibinfo{person}{Miao Hu} {et~al\mbox{.}}} \bibinfo{year}{2014}\natexlab{}.
\newblock \showarticletitle{Memristor crossbar-based neuromorphic computing system: A case study}.
\newblock \bibinfo{journal}{\emph{IEEE transactions on neural networks and learning systems}} \bibinfo{volume}{25}, \bibinfo{number}{10} (\bibinfo{year}{2014}), \bibinfo{pages}{1864--1878}.
\newblock


\bibitem[Kaggle(2012)]%
        {kddcup}
\bibfield{author}{\bibinfo{person}{Kaggle}.} \bibinfo{year}{2012}\natexlab{}.
\newblock \bibinfo{title}{KDD Cup 2012 Track2}.
\newblock \bibinfo{howpublished}{\url{https://www.kaggle.com/c/kddcup2012-track2/data}}.
\newblock


\bibitem[Kaggle(2014a)]%
        {avazu}
\bibfield{author}{\bibinfo{person}{Kaggle}.} \bibinfo{year}{2014}\natexlab{a}.
\newblock \bibinfo{title}{Avazu CTR Prediction}.
\newblock \bibinfo{howpublished}{\url{https://www.kaggle.com/c/avazu-ctr-prediction/data}}.
\newblock


\bibitem[Kaggle(2014b)]%
        {criteo}
\bibfield{author}{\bibinfo{person}{Kaggle}.} \bibinfo{year}{2014}\natexlab{b}.
\newblock \bibinfo{title}{Criteo Display Ad Challenge}.
\newblock \bibinfo{howpublished}{\url{https://www.kaggle.com/c/criteo-display-ad-challenge}}.
\newblock


\bibitem[Ke et~al\mbox{.}(2019)]%
        {ke2019recnmp}
\bibfield{author}{\bibinfo{person}{Liu Ke}, \bibinfo{person}{Udit Gupta}, \bibinfo{person}{Carole-Jean Wu}, \bibinfo{person}{Benjamin~Youngjae Cho}, \bibinfo{person}{Mark Hempstead}, \bibinfo{person}{Brandon Reagen}, \bibinfo{person}{Xuan Zhang}, \bibinfo{person}{David Brooks}, \bibinfo{person}{Vikas Chandra}, \bibinfo{person}{Utku Diril}, \bibinfo{person}{Amin Firoozshahian}, \bibinfo{person}{Kim Hazelwood}, \bibinfo{person}{Bill Jia}, \bibinfo{person}{Hsien-Hsin~S. Lee}, \bibinfo{person}{Meng Li}, \bibinfo{person}{Bert Maher}, \bibinfo{person}{Dheevatsa Mudigere}, \bibinfo{person}{Maxim Naumov}, \bibinfo{person}{Martin Schatz}, \bibinfo{person}{Mikhail Smelyanskiy}, {and} \bibinfo{person}{Xiaodong Wang}.} \bibinfo{year}{2019}\natexlab{}.
\newblock \bibinfo{title}{RecNMP: Accelerating Personalized Recommendation with Near-Memory Processing}.
\newblock
\newblock
\showeprint[arxiv]{1912.12953}~[cs.DC]


\bibitem[Li et~al\mbox{.}(2022)]%
        {li2022autolossgen}
\bibfield{author}{\bibinfo{person}{Zelong Li}, \bibinfo{person}{Jianchao Ji}, \bibinfo{person}{Yingqiang Ge}, {and} \bibinfo{person}{Yongfeng Zhang}.} \bibinfo{year}{2022}\natexlab{}.
\newblock \showarticletitle{AutoLossGen: Automatic Loss Function Generation for Recommender Systems}. In \bibinfo{booktitle}{\emph{Proceedings of the 45th International ACM SIGIR Conference on Research and Development in Information Retrieval}} (Madrid, Spain) \emph{(\bibinfo{series}{SIGIR '22})}. \bibinfo{publisher}{Association for Computing Machinery}, \bibinfo{address}{New York, NY, USA}, \bibinfo{pages}{1304–1315}.
\newblock
\showISBNx{9781450387323}
\urldef\tempurl%
\url{https://doi.org/10.1145/3477495.3531941}
\showDOI{\tempurl}


\bibitem[Lian et~al\mbox{.}(2018)]%
        {lian2018xdeepfm}
\bibfield{author}{\bibinfo{person}{Jianxun Lian}, \bibinfo{person}{Xiaohuan Zhou}, \bibinfo{person}{Fuzheng Zhang}, \bibinfo{person}{Zhongxia Chen}, \bibinfo{person}{Xing Xie}, {and} \bibinfo{person}{Guangzhong Sun}.} \bibinfo{year}{2018}\natexlab{}.
\newblock \showarticletitle{xDeepFM: Combining Explicit and Implicit Feature Interactions for Recommender Systems}. In \bibinfo{booktitle}{\emph{Proceedings of the 24th ACM SIGKDD International Conference on Knowledge Discovery \& Data Mining}} (London, United Kingdom) \emph{(\bibinfo{series}{KDD '18})}. \bibinfo{publisher}{Association for Computing Machinery}, \bibinfo{address}{New York, NY, USA}, \bibinfo{pages}{1754–1763}.
\newblock
\showISBNx{9781450355520}
\urldef\tempurl%
\url{https://doi.org/10.1145/3219819.3220023}
\showDOI{\tempurl}


\bibitem[Lin et~al\mbox{.}(2022)]%
        {lin2021naas}
\bibfield{author}{\bibinfo{person}{Yujun Lin}, \bibinfo{person}{Mengtian Yang}, {and} \bibinfo{person}{Song Han}.} \bibinfo{year}{2022}\natexlab{}.
\newblock \showarticletitle{NAAS: Neural Accelerator Architecture Search}. In \bibinfo{booktitle}{\emph{Proceedings of the 58th Annual ACM/IEEE Design Automation Conference}} (San Francisco, California, United States) \emph{(\bibinfo{series}{DAC '21})}. \bibinfo{publisher}{IEEE Press}, \bibinfo{pages}{1051–1056}.
\newblock
\showISBNx{9781665432740}
\urldef\tempurl%
\url{https://doi.org/10.1109/DAC18074.2021.9586250}
\showDOI{\tempurl}


\bibitem[Liu et~al\mbox{.}(2020)]%
        {liu2020autofis}
\bibfield{author}{\bibinfo{person}{Bin Liu}, \bibinfo{person}{Chenxu Zhu}, \bibinfo{person}{Guilin Li}, \bibinfo{person}{Weinan Zhang}, \bibinfo{person}{Jincai Lai}, \bibinfo{person}{Ruiming Tang}, \bibinfo{person}{Xiuqiang He}, \bibinfo{person}{Zhenguo Li}, {and} \bibinfo{person}{Yong Yu}.} \bibinfo{year}{2020}\natexlab{}.
\newblock \showarticletitle{AutoFIS: Automatic Feature Interaction Selection in Factorization Models for Click-Through Rate Prediction}. In \bibinfo{booktitle}{\emph{Proceedings of the 26th ACM SIGKDD International Conference on Knowledge Discovery \& Data Mining}} (Virtual Event, CA, USA) \emph{(\bibinfo{series}{KDD '20})}. \bibinfo{publisher}{Association for Computing Machinery}, \bibinfo{address}{New York, NY, USA}, \bibinfo{pages}{2636–2645}.
\newblock
\showISBNx{9781450379984}
\urldef\tempurl%
\url{https://doi.org/10.1145/3394486.3403314}
\showDOI{\tempurl}


\bibitem[Liu et~al\mbox{.}(2018)]%
        {liu2018darts}
\bibfield{author}{\bibinfo{person}{Hanxiao Liu}, \bibinfo{person}{Karen Simonyan}, {and} \bibinfo{person}{Yiming Yang}.} \bibinfo{year}{2018}\natexlab{}.
\newblock \showarticletitle{Darts: Differentiable architecture search}.
\newblock \bibinfo{journal}{\emph{arXiv preprint arXiv:1806.09055}} (\bibinfo{year}{2018}).
\newblock


\bibitem[Naumov et~al\mbox{.}(2019)]%
        {naumov2019deep}
\bibfield{author}{\bibinfo{person}{Maxim Naumov} {et~al\mbox{.}}} \bibinfo{year}{2019}\natexlab{}.
\newblock \showarticletitle{Deep learning recommendation model for personalization and recommendation systems}.
\newblock \bibinfo{journal}{\emph{arXiv preprint arXiv:1906.00091}} (\bibinfo{year}{2019}).
\newblock


\bibitem[Rendle(2012)]%
        {rendle2012factorization}
\bibfield{author}{\bibinfo{person}{Steffen Rendle}.} \bibinfo{year}{2012}\natexlab{}.
\newblock \showarticletitle{Factorization machines with libfm}.
\newblock \bibinfo{journal}{\emph{ACM Transactions on Intelligent Systems and Technology (TIST)}} \bibinfo{volume}{3}, \bibinfo{number}{3} (\bibinfo{year}{2012}), \bibinfo{pages}{1--22}.
\newblock


\bibitem[Sheng et~al\mbox{.}(2021)]%
        {sheng2021one}
\bibfield{author}{\bibinfo{person}{Xiang-Rong Sheng}, \bibinfo{person}{Liqin Zhao}, \bibinfo{person}{Guorui Zhou}, \bibinfo{person}{Xinyao Ding}, \bibinfo{person}{Binding Dai}, \bibinfo{person}{Qiang Luo}, \bibinfo{person}{Siran Yang}, \bibinfo{person}{Jingshan Lv}, \bibinfo{person}{Chi Zhang}, \bibinfo{person}{Hongbo Deng}, {and} \bibinfo{person}{Xiaoqiang Zhu}.} \bibinfo{year}{2021}\natexlab{}.
\newblock \showarticletitle{One Model to Serve All: Star Topology Adaptive Recommender for Multi-Domain CTR Prediction}. In \bibinfo{booktitle}{\emph{Proceedings of the 30th ACM International Conference on Information \& Knowledge Management}} (Virtual Event, Queensland, Australia) \emph{(\bibinfo{series}{CIKM '21})}. \bibinfo{publisher}{Association for Computing Machinery}, \bibinfo{address}{New York, NY, USA}, \bibinfo{pages}{4104–4113}.
\newblock
\showISBNx{9781450384469}
\urldef\tempurl%
\url{https://doi.org/10.1145/3459637.3481941}
\showDOI{\tempurl}


\bibitem[Song et~al\mbox{.}(2020)]%
        {song2020towards}
\bibfield{author}{\bibinfo{person}{Qingquan Song} {et~al\mbox{.}}} \bibinfo{year}{2020}\natexlab{}.
\newblock \showarticletitle{Towards automated neural interaction discovery for click-through rate prediction}. In \bibinfo{booktitle}{\emph{Proceedings of the 26th ACM SIGKDD International Conference on Knowledge Discovery \& Data Mining}}. \bibinfo{pages}{945--955}.
\newblock


\bibitem[Song et~al\mbox{.}(2019)]%
        {song2019autoint}
\bibfield{author}{\bibinfo{person}{Weiping Song}, \bibinfo{person}{Chence Shi}, \bibinfo{person}{Zhiping Xiao}, \bibinfo{person}{Zhijian Duan}, \bibinfo{person}{Yewen Xu}, \bibinfo{person}{Ming Zhang}, {and} \bibinfo{person}{Jian Tang}.} \bibinfo{year}{2019}\natexlab{}.
\newblock \showarticletitle{AutoInt: Automatic Feature Interaction Learning via Self-Attentive Neural Networks}. In \bibinfo{booktitle}{\emph{Proceedings of the 28th ACM International Conference on Information and Knowledge Management}} (Beijing, China) \emph{(\bibinfo{series}{CIKM '19})}. \bibinfo{publisher}{Association for Computing Machinery}, \bibinfo{address}{New York, NY, USA}, \bibinfo{pages}{1161–1170}.
\newblock
\showISBNx{9781450369763}
\urldef\tempurl%
\url{https://doi.org/10.1145/3357384.3357925}
\showDOI{\tempurl}


\bibitem[Wan et~al\mbox{.}(2020)]%
        {transpose_xb}
\bibfield{author}{\bibinfo{person}{Weier Wan}, \bibinfo{person}{Rajkumar Kubendran}, \bibinfo{person}{S.~Burc Eryilmaz}, \bibinfo{person}{Wenqiang Zhang}, \bibinfo{person}{Yan Liao}, \bibinfo{person}{Dabin Wu}, \bibinfo{person}{Stephen Deiss}, \bibinfo{person}{Bin Gao}, \bibinfo{person}{Priyanka Raina}, \bibinfo{person}{Siddharth Joshi}, \bibinfo{person}{Huaqiang Wu}, \bibinfo{person}{Gert Cauwenberghs}, {and} \bibinfo{person}{H.-S.~Philip Wong}.} \bibinfo{year}{2020}\natexlab{}.
\newblock \showarticletitle{33.1 A 74 TMACS/W CMOS-RRAM Neurosynaptic Core with Dynamically Reconfigurable Dataflow and In-situ Transposable Weights for Probabilistic Graphical Models}. In \bibinfo{booktitle}{\emph{2020 IEEE International Solid-State Circuits Conference - (ISSCC)}}. \bibinfo{pages}{498--500}.
\newblock
\urldef\tempurl%
\url{https://doi.org/10.1109/ISSCC19947.2020.9062979}
\showDOI{\tempurl}


\bibitem[Wang et~al\mbox{.}(2021a)]%
        {wang2021dcn}
\bibfield{author}{\bibinfo{person}{Ruoxi Wang} {et~al\mbox{.}}} \bibinfo{year}{2021}\natexlab{a}.
\newblock \showarticletitle{Dcn v2: Improved deep \& cross network and practical lessons for web-scale learning to rank systems}. In \bibinfo{booktitle}{\emph{Proceedings of the web conference 2021}}. \bibinfo{pages}{1785--1797}.
\newblock


\bibitem[Wang et~al\mbox{.}(2021b)]%
        {wang2021rerec}
\bibfield{author}{\bibinfo{person}{Yitu Wang} {et~al\mbox{.}}} \bibinfo{year}{2021}\natexlab{b}.
\newblock \showarticletitle{Rerec: In-reram acceleration with access-aware mapping for personalized recommendation}. In \bibinfo{booktitle}{\emph{2021 IEEE/ACM International Conference On Computer Aided Design (ICCAD)}}. IEEE, \bibinfo{pages}{1--9}.
\newblock


\bibitem[Wu et~al\mbox{.}(2024)]%
        {bwq}
\bibfield{author}{\bibinfo{person}{Xueying Wu}, \bibinfo{person}{Edward Hanson}, \bibinfo{person}{Nansu Wang}, \bibinfo{person}{Qilin Zheng}, \bibinfo{person}{Xiaoxuan Yang}, \bibinfo{person}{Huanrui Yang}, \bibinfo{person}{Shiyu Li}, \bibinfo{person}{Feng Cheng}, \bibinfo{person}{Partha~Pratim Pande}, \bibinfo{person}{Janardhan~Rao Doppa}, {et~al\mbox{.}}} \bibinfo{year}{2024}\natexlab{}.
\newblock \showarticletitle{Block-Wise Mixed-Precision Quantization: Enabling High Efficiency for Practical ReRAM-based DNN Accelerators}.
\newblock \bibinfo{journal}{\emph{IEEE Transactions on Computer-Aided Design of Integrated Circuits and Systems}} (\bibinfo{year}{2024}).
\newblock


\bibitem[Yan et~al\mbox{.}(2022)]%
        {sramimc}
\bibfield{author}{\bibinfo{person}{Bonan Yan}, \bibinfo{person}{Jeng-Long Hsu}, \bibinfo{person}{Pang-Cheng Yu}, \bibinfo{person}{Chia-Chi Lee}, \bibinfo{person}{Yaojun Zhang}, \bibinfo{person}{Wenshuo Yue}, \bibinfo{person}{Guoqiang Mei}, \bibinfo{person}{Yuchao Yang}, \bibinfo{person}{Yue Yang}, \bibinfo{person}{Hai Li}, \bibinfo{person}{Yiran Chen}, {and} \bibinfo{person}{Ru Huang}.} \bibinfo{year}{2022}\natexlab{}.
\newblock \showarticletitle{A 1.041-Mb/mm2 27.38-TOPS/W Signed-INT8 Dynamic-Logic-Based ADC-less SRAM Compute-in-Memory Macro in 28nm with Reconfigurable Bitwise Operation for AI and Embedded Applications}. In \bibinfo{booktitle}{\emph{2022 IEEE International Solid-State Circuits Conference (ISSCC)}}, Vol.~\bibinfo{volume}{65}. \bibinfo{pages}{188--190}.
\newblock
\urldef\tempurl%
\url{https://doi.org/10.1109/ISSCC42614.2022.9731545}
\showDOI{\tempurl}


\bibitem[Yang et~al\mbox{.}(2023)]%
        {yang2023pimpr}
\bibfield{author}{\bibinfo{person}{Tao Yang} {et~al\mbox{.}}} \bibinfo{year}{2023}\natexlab{}.
\newblock \showarticletitle{PIMPR: PIM-based Personalized Recommendation with Heterogeneous Memory Hierarchy}. In \bibinfo{booktitle}{\emph{2023 Design, Automation \& Test in Europe Conference \& Exhibition (DATE)}}. IEEE, \bibinfo{pages}{1--6}.
\newblock


\bibitem[Yang et~al\mbox{.}(2021)]%
        {yang2021multi}
\bibfield{author}{\bibinfo{person}{Xiaoxuan Yang} {et~al\mbox{.}}} \bibinfo{year}{2021}\natexlab{}.
\newblock \showarticletitle{Multi-objective optimization of ReRAM crossbars for robust DNN inferencing under stochastic noise}. In \bibinfo{booktitle}{\emph{2021 IEEE/ACM International Conference On Computer Aided Design (ICCAD)}}. IEEE, \bibinfo{pages}{1--9}.
\newblock


\bibitem[Yang et~al\mbox{.}(2022)]%
        {yang2022research}
\bibfield{author}{\bibinfo{person}{Xiaoxuan Yang} {et~al\mbox{.}}} \bibinfo{year}{2022}\natexlab{}.
\newblock \showarticletitle{Research progress on memristor: From synapses to computing systems}.
\newblock \bibinfo{journal}{\emph{IEEE Transactions on Circuits and Systems I: Regular Papers}} \bibinfo{volume}{69}, \bibinfo{number}{5} (\bibinfo{year}{2022}), \bibinfo{pages}{1845--1857}.
\newblock


\bibitem[Yang et~al\mbox{.}(2020)]%
        {retrans}
\bibfield{author}{\bibinfo{person}{Xiaoxuan Yang}, \bibinfo{person}{Bonan Yan}, \bibinfo{person}{Hai Li}, {and} \bibinfo{person}{Yiran Chen}.} \bibinfo{year}{2020}\natexlab{}.
\newblock \showarticletitle{ReTransformer: ReRAM-based processing-in-memory architecture for transformer acceleration}. In \bibinfo{booktitle}{\emph{Proceedings of the 39th International Conference on Computer-Aided Design}} (Virtual Event, USA) \emph{(\bibinfo{series}{ICCAD '20})}. \bibinfo{publisher}{Association for Computing Machinery}, \bibinfo{address}{New York, NY, USA}, Article \bibinfo{articleno}{92}, \bibinfo{numpages}{9}~pages.
\newblock
\showISBNx{9781450380263}
\urldef\tempurl%
\url{https://doi.org/10.1145/3400302.3415640}
\showDOI{\tempurl}


\bibitem[Yu et~al\mbox{.}(2020)]%
        {yu2020bignas}
\bibfield{author}{\bibinfo{person}{Jiahui Yu} {et~al\mbox{.}}} \bibinfo{year}{2020}\natexlab{}.
\newblock \showarticletitle{Bignas: Scaling up neural architecture search with big single-stage models}. In \bibinfo{booktitle}{\emph{Computer Vision--ECCV 2020: 16th European Conference, Glasgow, UK, August 23--28, 2020, Proceedings, Part VII 16}}. Springer, \bibinfo{pages}{702--717}.
\newblock


\bibitem[Zhang et~al\mbox{.}(2023a)]%
        {zhang2023distdnas}
\bibfield{author}{\bibinfo{person}{Tunhou Zhang} {et~al\mbox{.}}} \bibinfo{year}{2023}\natexlab{a}.
\newblock \showarticletitle{DistDNAS: Search Efficient Feature Interactions within 2 Hours}.
\newblock \bibinfo{journal}{\emph{arXiv preprint arXiv:2311.00231}} (\bibinfo{year}{2023}).
\newblock


\bibitem[Zhang et~al\mbox{.}(2023b)]%
        {zhang2023NASRec}
\bibfield{author}{\bibinfo{person}{Tunhou Zhang} {et~al\mbox{.}}} \bibinfo{year}{2023}\natexlab{b}.
\newblock \showarticletitle{NASRec: weight sharing neural architecture search for recommender systems}. In \bibinfo{booktitle}{\emph{Proceedings of the ACM Web Conference 2023}}. \bibinfo{pages}{1199--1207}.
\newblock


\bibitem[Zhang et~al\mbox{.}(2025)]%
        {zhang2023nasrec2}
\bibfield{author}{\bibinfo{person}{Tunhou Zhang}, \bibinfo{person}{Dehua Cheng}, \bibinfo{person}{Yuchen He}, \bibinfo{person}{Zhengxing Chen}, \bibinfo{person}{Xiaoliang Dai}, \bibinfo{person}{Liang Xiong}, \bibinfo{person}{Yudong Liu}, \bibinfo{person}{Feng Cheng}, \bibinfo{person}{Yufan Cao}, \bibinfo{person}{Feng Yan}, {et~al\mbox{.}}} \bibinfo{year}{2025}\natexlab{}.
\newblock \showarticletitle{Towards Automated Model Design on Recommender Systems}.
\newblock \bibinfo{journal}{\emph{ACM Transactions on Recommender Systems}} \bibinfo{volume}{3}, \bibinfo{number}{3} (\bibinfo{year}{2025}), \bibinfo{pages}{1--23}.
\newblock


\bibitem[Zhaok et~al\mbox{.}(2021)]%
        {zhaok2021autoemb}
\bibfield{author}{\bibinfo{person}{Xiangyu Zhaok} {et~al\mbox{.}}} \bibinfo{year}{2021}\natexlab{}.
\newblock \showarticletitle{Autoemb: Automated embedding dimensionality search in streaming recommendations}. In \bibinfo{booktitle}{\emph{2021 IEEE International Conference on Data Mining (ICDM)}}. IEEE, \bibinfo{pages}{896--905}.
\newblock


\bibitem[Zheng et~al\mbox{.}(2023)]%
        {mbsa}
\bibfield{author}{\bibinfo{person}{Qilin Zheng}, \bibinfo{person}{Shiyu Li}, \bibinfo{person}{Yitu Wang}, \bibinfo{person}{Ziru Li}, \bibinfo{person}{Yiran Chen}, {and} \bibinfo{person}{Hai~Helen Li}.} \bibinfo{year}{2023}\natexlab{}.
\newblock \showarticletitle{Accelerating Sparse Attention with a Reconfigurable Non-volatile Processing-In-Memory Architecture}. In \bibinfo{booktitle}{\emph{2023 60th ACM/IEEE Design Automation Conference (DAC)}}. \bibinfo{pages}{1--6}.
\newblock
\urldef\tempurl%
\url{https://doi.org/10.1109/DAC56929.2023.10247908}
\showDOI{\tempurl}


\bibitem[Zhu et~al\mbox{.}(2020)]%
        {mnsim}
\bibfield{author}{\bibinfo{person}{Zhenhua Zhu}, \bibinfo{person}{Hanbo Sun}, \bibinfo{person}{Kaizhong Qiu}, \bibinfo{person}{Lixue Xia}, \bibinfo{person}{Gokul Krishnan}, \bibinfo{person}{Guohao Dai}, \bibinfo{person}{Dimin Niu}, \bibinfo{person}{Xiaoming Chen}, \bibinfo{person}{Xiaobo~Sharon Hu}, \bibinfo{person}{Yu Cao}, \bibinfo{person}{Yuan Xie}, \bibinfo{person}{Yu Wang}, {and} \bibinfo{person}{Huazhong Yang}.} \bibinfo{year}{2020}\natexlab{}.
\newblock \showarticletitle{{MNSIM} 2.0: {A} Behavior-Level Modeling Tool for Memristor-based Neuromorphic Computing Systems}. In \bibinfo{booktitle}{\emph{{GLSVLSI} '20: Great Lakes Symposium on {VLSI} 2020, Virtual Event, China, September 7-9, 2020}}, \bibfield{editor}{\bibinfo{person}{Tinoosh Mohsenin}, \bibinfo{person}{Weisheng Zhao}, \bibinfo{person}{Yiran Chen}, {and} \bibinfo{person}{Onur Mutlu}} (Eds.). \bibinfo{publisher}{{ACM}}, \bibinfo{pages}{83--88}.
\newblock


\end{thebibliography}

\end{document}